\documentclass[11pt]{article}

\usepackage{epsfig}
\usepackage{subfig}
\usepackage{graphicx}
\usepackage{epstopdf}
\usepackage{amsfonts}
\usepackage{amssymb}
\usepackage{amsthm}
\usepackage{amsmath}
\usepackage{multirow}
\usepackage{color}
\usepackage{cite}

\usepackage{xcolor}
 \usepackage{makecell}
\def\beq{\begin{equation}}
\def\eeq{\end{equation}}
\def\bea{\begin{eqnarray}}
\def\eea{\end{eqnarray}}
\newcommand{\beqs}{\begin{subequations}}
\newcommand{\eeqs}{\end{subequations}}

\newcommand{\cref}[1]{Ref.~\cite{#1}}

\newcommand{\hh}{{\ensuremath{I{\kern-2.6pt h}}}}
\newcommand{\bhh}{{\ensuremath{\bar{I{\kern-2.6pt h}}}}}

\newcommand\blfootnote[1]{%
  \begingroup
  \renewcommand\thefootnote{}\footnote{#1}%
  \addtocounter{footnote}{-1}%
  \endgroup
}

\usepackage[colorlinks=true,allcolors=blue]{hyperref}

\begin{document}

\begin{titlepage}
	
%\vspace*{-15mm}
%\begin{flushright}
%{UT-STPD-21/01}\\
%\end{flushright}
%\vspace*{0.7cm}

\begin{center}
{\Large {\bf Gravitational waves from walls bounded by strings in $SO(10)$ model of pseudo-Goldstone dark matter} \blfootnote{\color{blue}All authors contributed equally.}
}
\\[12mm]
%George Lazarides,$^{1}$
Rinku Maji,$^{1}$
Wan-Il Park,$^{2,3,4}$
Qaisar Shafi$^{5}$~%\footnote{E-mail: \texttt{shafi@bartol.udel.edu}}
\end{center}
\vspace*{0.50cm}
	\centerline{$^{1}$ \it
		Laboratory for Symmetry and Structure of the Universe, Department of Physics,}
		\centerline{\it  Jeonbuk National University, Jeonju 54896, Republic of Korea}
	\vspace*{0.2cm}
	\centerline{$^{2}$ \it
		Division of Science Education and Institute of Fusion Science,}
		\centerline{\it  Jeonbuk National University, Jeonju 54896, Republic of Korea}
	\vspace*{0.2cm}
	\centerline{$^{3}$ \it
		Instituto de Física Corpuscular,}
		\centerline{\it CSIC-Universitat de València, C/Catedrático José Beltrán 2, Paterna 46980, Spain}
	\vspace*{0.2cm}
	\centerline{$^{4}$ \it
		Departament de Física Teòrica,}
		\centerline{\it Universitat de València, C/ Dr. Moliner 50, Burjassot 46100, Spain}			
	\vspace*{0.2cm}
	\centerline{$^{5}$ \it
		Bartol Research Institute, Department of Physics and 
		Astronomy,}
	\centerline{\it
		 University of Delaware, Newark, DE 19716, USA}
	\vspace*{1.20cm}
\begin{abstract}
We explore the gravitational wave spectrum generated by string-wall structures in an $SO(10)$ ($Spin(10)$) based scenario of pseudo-Goldstone boson dark matter (pGDM) particle. This dark matter candidate is a linear combination of the Standard Model (SM) singlets present in the 126 and 16 dimensional Higgs fields. The Higgs $126$-plet vacuum expectation value (VEV) $\left<126_H\right>$ leaves unbroken the $\mathbb{Z}_2$ subgroup of $\mathbb{Z}_4$, the center of $SO(10)$. Among other things, this yields topologically stable cosmic strings with a string tension $\mu \sim \left<126_H\right>^2$. The subsequent (spontaneous) breaking of $\mathbb{Z}_2$ at a significantly lower scale by the $16$-plet VEV $\left<16_H\right>$ leads to the appearance of domain walls bounded by the strings produced earlier. We display the gravitational wave spectrum for $G \mu$ values varying between $10^{-15}$ and $10^{-9}$ ($\left<126_H\right>\sim 10^{11}$ - $10^{14}$ GeV), and $\left<16_H\right>\sim 0.1$ - $10^2$ TeV range ($G$ denotes Newton’s constant.) These predictions can be tested, as we show, by a variety of (proposed) experiments including LISA, ET, CE and others.

\end{abstract}

\end{titlepage}
%%%%%%%%%%%%%%%%%%%%%%%%%%%%%%%%%%%%%%%%%%%%%%%%%%%%%%%
\section{Introduction}

﻿It is widely agreed that the Standard Model (SM) does not contain a plausible dark matter candidate which makes up close to $25\%$ of the universe’s energy density \cite{Planck:2018vyg}. A number of direct detection experiments have imposed fairly stringent constraints on a variety of WIMP-like dark matter candidates \cite{PandaX-4T:2021bab,LZ:2022ufs,XENON:2023sxq}. A pseudo-Goldstone Dark Matter (pGDM) model has attracted a fair amount of attention in recent years with the characteristic feature that the dark matter candidate is able to evade the direct detection constraints \cite{Gross:2017dan,Alanne:2018zjm,Huitu:2018gbc,Karamitros:2019ewv,Jiang:2019soj,Abe:2020iph,Okada:2020zxo,Abe:2020ldj,Zhang:2021alu,Abe:2021byq,Okada:2021qmi,Mohapatra:2021ozu,Liu:2022evb,Jiang:2023xdf,Mohapatra:2023aei,Qiu:2023wbs}. Realistic examples implementing this scenario are based on well-known local gauge symmetries including $U(1)_{B-L}$ \cite{Abe:2020iph,Okada:2020zxo} and $SO(10)$ \cite{Abe:2021byq,Okada:2021qmi}.

In the $SO(10)$ model the dark matter candidate is a pseudo-Goldstone particle formed from a suitable linear combination of the SM singlets contained in the Higgs 126-plet and 16-plet fields. We briefly summarize here the salient features of this model.
\begin{itemize}
\item[1.] The dark matter can decay via gauge interactions. The VEV along the SM singlet direction in $126_H$ should be above $10^{11}$ GeV \cite{Abe:2021byq,Okada:2021qmi} in order to satisfy the lifetime bound, $\tau_{\rm DM}\gtrsim 10^{27}$ sec \cite{Baring:2015sza}, for decaying dark matter. The right-handed neutrinos ($\nu^C_L$) acquire Majorana masses from this VEV and it is worth noting that the above requirement $\left<126_H\right>\gtrsim 10^{11}$ GeV from DM considerations coincides with the requirement of the right-handed neutrino masses preferred by the fitting of neutrino data and successful leptogenesis in $SO(10)$ GUT \cite{Lazarides:1980nt, Babu:1992ia, Bajc:2005zf, Joshipura:2011nn, Altarelli:2013aqa, Dueck:2013gca, Meloni:2014rga, Meloni:2016rnt,Babu:2016bmy,Ohlsson:2018qpt, Boucenna:2018wjc, Ohlsson:2019sja, Mummidi:2021anm,Lazarides:2022ezc}.

\item[2.] The VEV of $16_H$ should be of the order of the electroweak scale ($\sim 10^2 - 10^3$ GeV) for a viable pNGB thermal dark matter. However, near the resonance region, it could increase to $\sim 10^5$ GeV or so, namely, $\left<16_H\right>/v_H\simeq [1,10^3]$, with $v_H$ being the VEV of the SM Higgs \cite{Abe:2020iph,Okada:2020zxo,Abe:2021byq,Okada:2021qmi}.
\end{itemize} 

 The $126_H$ and $16_H$ VEVs generate composite topological structures known as domain walls bounded by strings with a well-defined hierarchy between them. These topological defects produce stochastic gravitational wave background \cite{Martin:1996ea,Dunsky:2021tih}, which could be detected in many proposed gravitational wave experiments, including LISA \cite{Bartolo:2016ami, amaroseoane2017laser}, DECIGO \cite{Sato_2017}, BBO \cite{Crowder:2005nr, Corbin:2005ny}, CE \cite{PhysRevLett.118.151105}, ET \cite{Mentasti:2020yyd}, and HLVK \cite{KAGRA:2013rdx}. 
 
 In Sec.~\ref{sec:model-TD}, we provide a brief outline of the model and the formation of walls bounded by strings. Sec.~\ref{sec:so10} discusses the generation of the stochastic gravitational wave background from the string-wall system and their observational prospects. Our conclusions are summarized in Sec.~\ref{sec:summary}.
%%%%%%%%%%%%%%%%%%%%%%%%%%%%%%%%%%%%%%%%%%%%%%%%%%%%%%
\section{Pseudo Goldstone dark matter model and walls bounded by strings}
\label{sec:model-TD}
 The scalar sector of the minimal pGDM model contains an SM-singlet complex scalar $S$ in addition to the SM Higgs doublet $H$. The scalar potential is given by \cite{Gross:2017dan},
\begin{align}\label{eq:pot-min}
V =- \mu_h^2 |H|^2 
- \mu_s^2 |S|^2 
+\lambda_h|H|^4 
+\lambda_{hs} |H|^2 |S|^2 
+ \lambda_s |S|^4 
- {\mu_s^\prime}^2  (S^2 + \mathrm{h.c.})
\end{align}
The Lagrangian possesses a $\mathbb{Z}_2$ symmetry, the subgroup of a global $U(1)$ symmetry $S\to e^{i\alpha}S$ which is softly broken by the last term in Eq.~\eqref{eq:pot-min}. As the radial component of $S$ gets a VEV, its angular component will be a pseudo Nambu-Goldstone boson (pNGB) due the spontaneous and explicit breaking of the $U(1)$ symmetry, and becomes a viable dark matter candidate stabilized by the CP symmetry ($S\to S^*$) \cite{Gross:2017dan}. The right amount of DM relic density can be obtained for the DM mass $m_\chi\approx [m_h/2,10~\mathrm{TeV}]$, where $m_h$ is the SM Higgs mass. There are two resonances around $m_\chi=m_h/2$ and $m_\chi=m_{h_2}/2$, where $m_{h_2}$ is the mass of the second BSM Higgs which comes from a linear combination of CP-even states. The direct detection cross-section is vanishingly small in the limit of zero momentum transfer due to its pseudo-Goldstone nature, which can alleviate the direct detection bounds \cite{PandaX-4T:2021bab,LZ:2022ufs,XENON:2023sxq} on WIMP-like dark matter. However, the breaking of $\mathbb{Z}_2$ generates stable domain walls which contradict the standard cosmology \cite{Zeldovich:1974uw}.

 An ultra-violate (UV) completion of the minimal model with a gauge $U(1)_{B-L}$ was proposed in Refs.~\cite{Abe:2020iph,Okada:2020zxo} with complex scalars $S$ and $\Phi$ carrying one and two units of $B-L$ charges respectively. In this model, the spontaneous breaking of $U(1)_{B-L}$ gauge symmetry by the $\Phi$ VEV leaves a remnant gauge $\mathbb{Z}_2$ symmetry and generates the soft breaking term in Eq.~\eqref{eq:pot-min} from the trilinear term 
\begin{align}
\beta( \Phi^\dagger S^2 + \mathrm{h.c.}) .
\end{align}
 A unified approach to implement these ideas is based on $SO(10)$ with $S\in 16_H$ and $\Phi\in \overline{126}_H$, and the trilinear term arising from the coupling ($\overline{126}_H(16_H)^2+$ h.c.) in the scalar potential.
%%%%%%%%%%%%%%%%%%%%%%%%%%%%%%%%%%%%%%%%%%%%%%%%%%%%%%

 This realistic $SO(10)$ model of pGDM \cite{Abe:2021byq,Okada:2021qmi} includes the electroweak Higgs doublet coming from a linear combination of bi-doublets in $\overline{126}_H$ and a complex $10_H$. The Yukawa couplings of the fermion $16_F$ with $\overline{126}_H$ and $10_H$ to produce realistic fermion masses have been extensively studied in the literature \cite{Lazarides:1980nt, Babu:1992ia, Bajc:2005zf, Joshipura:2011nn, Altarelli:2013aqa, Dueck:2013gca, Meloni:2014rga, Meloni:2016rnt,Babu:2016bmy,Ohlsson:2018qpt, Boucenna:2018wjc, Ohlsson:2019sja, Mummidi:2021anm,Lazarides:2022ezc,Saad:2022mzu}. A VEV along the SM singlet direction of $126_H$ or $16_H$ breaks a diagonal generator orthogonal to the hypercharge ($Y$) and reduces the rank of the gauge symmetry from five to four. An example is the symmetry breaking of $SU(2)_L\times SU(2)_R\times U(1)_{B-L}$ to $SU(2)_L\times U(1)_Y$, where $T_{3R}-\frac{B-L}{2}$ is broken. The dark matter candidate is the pseudo-Goldstone mode coming from a linear combination of the CP-odd components of the SM singlets. The gauge boson associated with the broken generator orthogonal to the hypercharge absorbs the massless would-be Goldstone mode.

The VEV of the $SU(5)$ singlet component in the scalar multiplet $\overline{126}_H$ ($\left<126_H\right>\gtrsim 10^{11}$ GeV) leaves an unbroken $\mathbb{Z}_2$ and therefore generates topologically stable cosmic strings \cite{Kibble:1982ae}. Subsequently, however, the VEV $\left<16_H\right>$ in the range $[10^2,10^5]$ GeV for the right amount of DM relic,  breaks this $\mathbb{Z}_2$ symmetry, which leads to the formation of domain walls bounded by strings \cite{Kibble:1982dd,Lazarides:1982tw,Lazarides:2023iim}.
These walls bounded by strings are distinct, of course, from those \cite{Kibble:1982dd} arising from the breaking of C- or  D-parity even intermediate Pati-Salam \cite{Pati:1974yy} or left-right symmetric gauge symmetry achieved by the VEV of $(1,1,1)\in 54_H$ or $(1,1,1,0)\in 210_H$ respectively. In these latter cases, the GUT-scale strings are produced along with topologically stable monopoles \cite{tHooft:1974kcl,Lazarides:1980cc,Kibble:1982ae}. The strings become the boundary of the domain walls generated during the subsequent breaking of this C-parity at an intermediate scale \cite{Kibble:1982dd}. In GUTs, the C-parity breaking scale is equal to or higher than the right-handed neutrino mass scale ($m_R\gtrsim 10^{11}$ GeV) \cite{Lazarides:1980nt}. Therefore, the proposed or ongoing experiments, sensitive upto kHz frequency, cannot observe the gravitational waves. The breaking of $SO(10)$  can accommodate one or multiple intermediate gauge symmetries depending on the choice of scalars $45_H$, $54_H$, or $210_H$, and the directions of the VEVs. The solution regions for the various breaking patterns have been extensively studied in the literature (see, for example, Refs.~\cite{Chakrabortty:2017mgi, Chakrabortty:2019fov} and the references therein).

%%%%%%%%%%%%%%%%%%%%%%%%%%%%%%%%%%%%%%%%%%%%%%%%%%%%%%%%%%%%%%%%%%%%%%
\section{Gravitational wave background}
\label{sec:so10}
%%%%%%%%%%%%%%%%%%%%%%%%%%%%%%%%%%%%%%%%%%%%%%%%%%%%%%%%%%%%%%%
In this section we discuss the gravitational waves from the domain walls bounded by strings and their observational prospects. The cosmic strings formed at $\left<126_H\right>$ have a tension (mass per unit length) given by \cite{Hill:1987qx,Hindmarsh:2011qj},
\begin{align}\label{eq:mu}
\mu\simeq \pi B(\frac{\lambda_{\rm str}}{g_{\rm str}^2})\left<126_H\right>^2 ,
\end{align} 
where $\lambda_{\rm str}$ and $g_{\rm str}$ are the relevant quartic and gauge coupling constants, and the function 
\begin{align}\label{eq:Bx}
B(x)\simeq \begin{cases} %1 & \mbox{for} \ x=1 \\
1.04 \ x^{0.195} & \mbox{for} \ 10^{-2}\lesssim x\lesssim 10^2 \\
2.4/\ln(2/x) & \mbox{for} \ x\lesssim 0.01.
\end{cases}
\end{align}
The tension on the domain walls (mass per unit area) associated with the breaking of $\mathbb{Z}_2$-symmetry by $\left<16_H\right>$ is given by, 
%\begin{align}\label{eq:sig-dw}
%\sigma = \frac{2\sqrt{2}}{3}\sqrt{\lambda_{\rm dw}} \left<16_H\right>^3 ,
%\end{align}
%with $\lambda_{\rm dw}$ being the quartic coupling of the relevant scalar.
\begin{align}\label{eq:sig-dw}
\sigma \simeq \xi_\chi \Delta V_\chi \simeq \xi_\chi m_\chi^2 \left<16_H\right>^2 \sim \sqrt{\lambda_{\rm dw}} \left<16_H\right>^3,
\end{align}
where $\xi_\chi$ is the correlation length of the pNGB field $\chi$, $\Delta V_\chi$ is the potential height along the direction, $m_\chi$ is the mass of the field, and $\lambda_{\rm dw}$ denotes the quartic coupling of the associated radial mode field. 
Note that in the last expression of Eq.~(\ref{eq:sig-dw}), considering the parameter space for the right amount of dark matter as a thermal relic, we restricted ourselves to the case in which $m_\chi$ is comparable to the mass of the radial mode.

%\textbf{\color{magenta}(When it is formed, the loop size is $\ell = \alpha t$ with $\alpha = 0.1$. Then, $R \approx \ell/ 2 \pi = \left( \alpha / 2 \pi \right) t$. ?)}
For a wall bounded by a string of radius of curvature $R$, the force per unit length on the string boundary $\sim \mu/R$ dominates over the wall tension $\sigma$ for $R<R_c=\mu/\sigma$. 
The maximum radius of curvature $R$ is of the order of the cosmic time $t$. 
Therefore, the string dynamics dominates before the formation of the domain walls ($t_{\rm dw}$) until time $R_c>t_{\rm dw}$. On the other hand, if $t_{\rm dw}>R_c$, the domain wall dynamics starts dominating right after their formation. We define the time $t_*=\mathrm{max}[R_c,t_{\rm dw}]$ to be the maximum timescale for domination by the string dynamics \cite{Vilenkin:1982ks,Vilenkin:1984ib,Martin:1996ea,Dunsky:2021tih}. The cosmic strings inter-commute, form loops before $t_*$ with $R<t_*$ and can produce gravitational waves \cite{Martins:1995tg,Martins:1996jp,Martin:1996ea,Martins:2000cs}. The domain wall dynamics become dominant for $t>t_*$, and the string-wall networks collapse as the walls pull the strings (see Ref.~\cite{Dunsky:2021tih} for a detailed analysis).

%%%%%%%%%%%%%%%%%%%%%%%%%%%%%%%%%%%%%%%%%%%%%%%%%%%%%%%%%%%
\begin{figure}
\begin{center}
\includegraphics[width=0.7\textwidth]{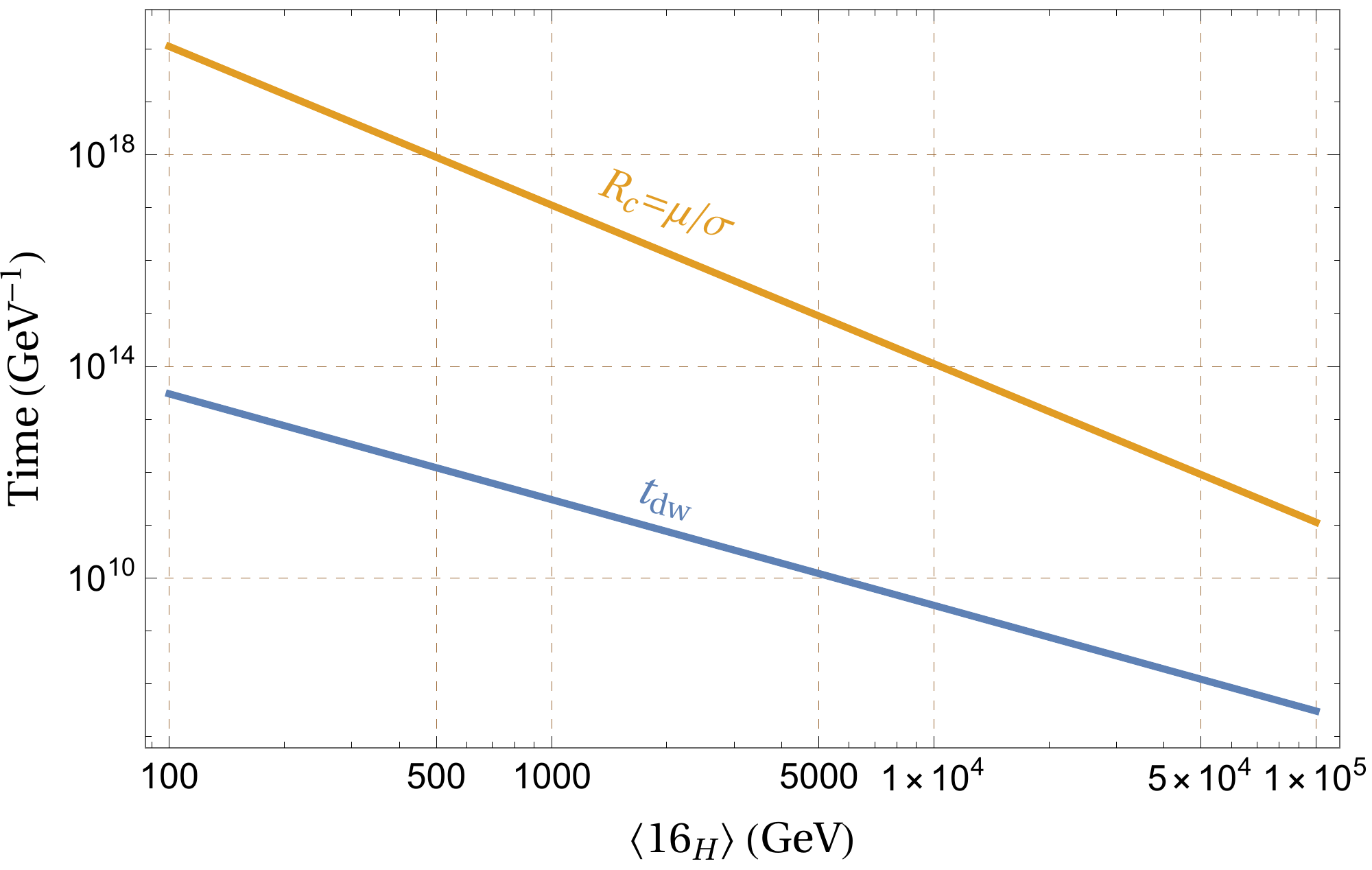}
\end{center}
\caption{Comparison of $R_c=\mu/\sigma$ and domain wall formation time $t_{\rm dw}$ as a function of the scale $\left<16_H\right>$ for a typical value $G\mu =10^{-12}$. We see that $R_c\gg t_{\rm dw}$ for this $SO(10)$ model.}\label{fig:tDW-Rc}
\end{figure}

%%%%%%%%%%%%%%%%%%%%%%%%%%%%%%%%%%%%%%%%%%%%%%%%%%%%%%%%%%%
The time for the domain wall formation $t_{\rm dw}$ in the radiation-dominated universe is given by
\begin{equation}\label{eq:time-rad-dom}
t_{\rm dw}=\sqrt{\frac{45}{2\pi^2g_*}}\,\frac{m_{\rm Pl}}{T_{\rm dw}^2} , 
\end{equation}
where $g_*$ accounts for the effective number of massless degrees of freedom, and we take $T_{\rm dw} \approx \left< 16_H \right>/\sqrt{\lambda_{\rm dw}}$ as the background temperature at the time of domain wall formation.
We have compared $R_c$ and $t_{\rm dw}$ in Fig.~\ref{fig:tDW-Rc} with $G\mu = 10^{-12}$ for the relevant range of the VEV $\left<16_H\right>\in[10^2,10^5]$ GeV.
Since $R_c$ is larger than the domain wall formation time $t_{\rm dw}$, the string-wall network starts collapsing at $t_*=R_c$ and loop formation ceases. After $t_* \approx R_c$, there will be  walls bounded by strings of curvature $R\lesssim R_c$ formed on or before $t_*$. The maximum radius of curvature of the wall bounded by string dominated by the wall dynamics could be $R_c$. The lifetime of such an object is $\tau_{\rm ws}\sim \pi (\Gamma G\sigma)^{-1}$, with $\Gamma\sim \mathcal{O}(10^2)$ being a numerical factor \cite{Vachaspati:1984gt,Dunsky:2021tih} which is not larger than the lifetime of decay $\tau_s\sim 2\pi(\Gamma G\sigma)^{-1}$ of a string loop with radius $R_c$. Moreover, the string-wall system oscillates relativistically after $R_c$ and can be chopped into smaller pieces making its lifetime much smaller.  Therefore, the contribution to the gravitational wave background will arise dominantly from the cosmic strings in the case of $t_{\rm dw}\ll R_c$ \cite{Martin:1996ea,Dunsky:2021tih}. 

%%%%%%%%%%%%%%%%%%%%%%%%%%%%%%%%%%%%%%%%%%%%%%%%%%%%%%%%%%%%
\begin{figure}[htbp]
\begin{center}
\includegraphics[width=0.9\textwidth]{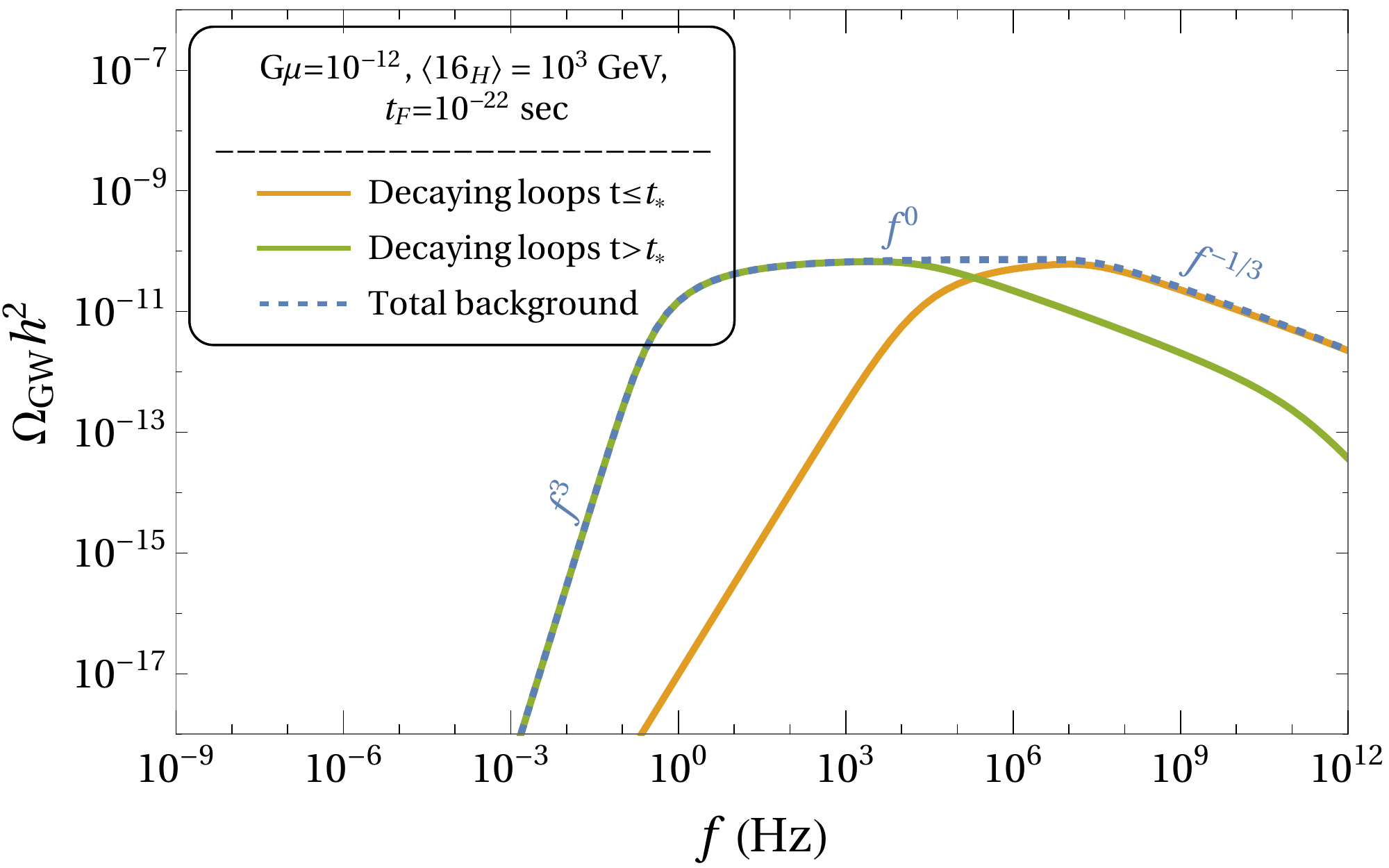}
\end{center}
\caption{Contribution to the gravitational wave background from decaying cosmic string loops at different era for a typical value of $G\mu=10^{-12}$ and $\left<16_H\right>=10^3$ GeV.}\label{fig:gws-comp}
\end{figure}
%%%%%%%%%%%%%%%%%%%%%%%%%%%%%%%%%%%%%%%%%%%%%%%%%%%%%%%%%%%%%%%%%

The gravitational wave background is  represented as a relic energy density as a function of frequency $f$ in the present time,
\begin{align}
\Omega_{\rm GW} = \frac{1}{\rho_c}\frac{d\rho_{\rm GW}}{d\log f}
\end{align}
 where $\rho_c$ is the critical energy density of the present universe. The stochastic gravitational wave background from domain walls bounded by strings is estimated in Ref.~\cite{Dunsky:2021tih} as a sum of the contributions from normal modes given by,
\begin{align}
   \label{eq:GWs1}
\Omega_{\rm GW}(f)=\sum_n\Omega^{(n)}_{\rm GW} ,
\end{align}
where
\begin{align}
   \label{eq:GWs2}
    \Omega^{(n)}_{\rm GW} 
    = \frac{1}{\rho_c} \int_{t_F}^{t_0} d\tilde{t} \left(\frac{a(\tilde{t})}{a(t_0)}\right)^5\frac{\mathcal{F} C_{\rm eff}(t_i)}{\alpha t_i^4} \left(\frac{a(t_i)}{a(\tilde{t})}\right)^3
     \frac{\left(1 + \frac{\xi n}{2\pi R_c f}\frac{a(\tilde{t})}{a(t_0)} \right)}{\Gamma G \mu + \alpha(1 + \frac{\alpha t_i}{2\pi R_c})}\frac{\Gamma n^{-q}}{\zeta(q)} G\mu^2 \frac{\xi n}{f}\theta(t_* - t_i) ,
\end{align}
and a loop of initial length $l_i=\alpha t_i$ decays with its length ($l$) at any subsequent time $t$ given by
\begin{align}
    \label{eq:lengthLoss}
    G\mu(t - t_i) = \int_l^{\alpha t_i}  dl' \frac{1 + \frac{l'}{2 \pi R_c}}{\Gamma(l')}.
\end{align}
 In our case, $t_* = R_c\gg t_{\rm dw}$, and therefore we have $\xi = 2$ and $\Gamma\simeq 50$ corresponding to the pure string limit \cite{Vachaspati:1984gt,Vilenkin:2000jqa}. We have taken $\mathcal{F} \simeq 0.1$, $\alpha\simeq 0.1$, $C_{\rm eff} = 5.7$ for the radiation-dominated universe \cite{Vanchurin:2005pa,Ringeval:2005kr,Olum:2006ix,Blanco-Pillado:2013qja,Blanco-Pillado:2017oxo,Cui:2018rwi}, and $q=4/3$ because of cusp domination in the gravitational wave spectrum \cite{Olmez:2010bi, Auclair:2019wcv, Cui:2019kkd, LIGOScientific:2021nrg}. We have included the sum of normal modes upto $n=10^7$.

Fig.~\ref{fig:gws-comp} shows the gravitational wave background for a typical value of $G\mu=10^{-12}$ and $\left<16_H\right>=10^3$ GeV.  There is a scale-invariant component ($f^0$), an $f^3$ power-law IR spectrum, and $f^{-1/3}$ UV tail which agrees with \cite{Cui:2019kkd,Dunsky:2021tih}. The loop formation ceases at $t_*$ and contribution to the gravitational wave background at time $t>t_*$ comes from the decaying loops formed before $t_*$. These decaying loops during $t>t_*$ contribute to the lower frequency region of the spectrum with an $f^3$ IR tail. The UV tail arises since we assume that the network appears at time $t_F=10^{-22}$ sec as an example, enters the scaling regime soon after $t_F$ and the earliest loops of size $\alpha t_F$ decay at time $\alpha t_F(\Gamma G\mu)^{-1}$.

%%%%%%%%%%%%%%%%%%%%%%%%%%%%%%%%%%%%%%%%%%%%%%%%%%%%%%%%%%%%%%%%%
\begin{figure}[htbp!]
\begin{center}
\includegraphics[width=0.9\textwidth]{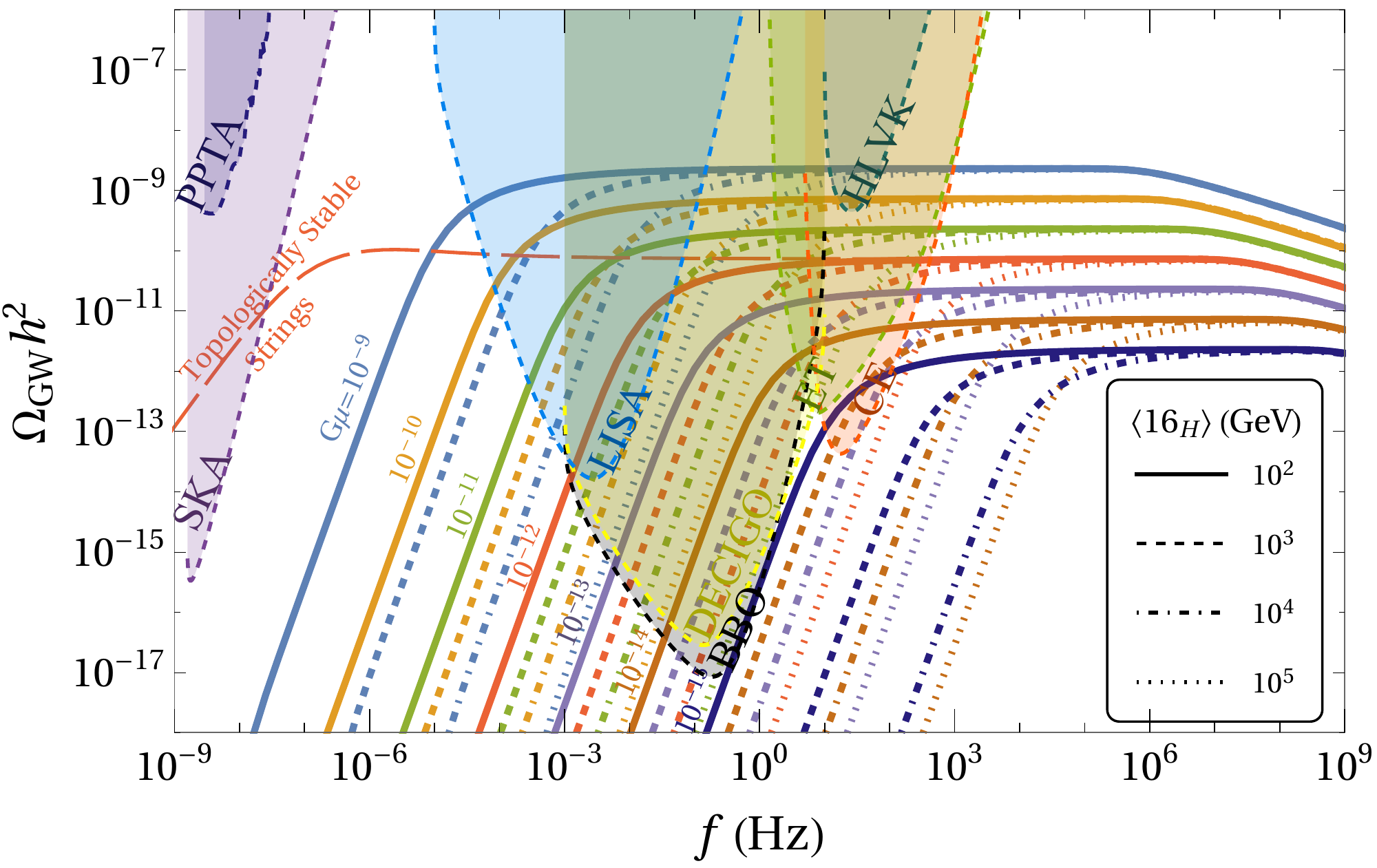}
\end{center}
\caption{Gravitational wave background from domain walls bounded by cosmic strings for $G\mu$ varying from $10^{-15}$ to $10^{-9}$, and the SM singlet VEV in $16_H$ varying from $10^2$ GeV to $10^5$ GeV. We have shown the sensitivity curves \cite{Thrane:2013oya, Schmitz:2020syl} for various ongoing and proposed experiments, namely, PPTA \cite{Shannon:2015ect}, SKA \cite{5136190, Janssen:2014dka}, CE \cite{PhysRevLett.118.151105}, ET \cite{Mentasti:2020yyd}, LISA \cite{Bartolo:2016ami, amaroseoane2017laser}, DECIGO \cite{Sato_2017}, BBO \cite{Crowder:2005nr, Corbin:2005ny} and HLVK \cite{KAGRA:2013rdx}. For comparison, we also display the gravitational wave background from topologically stable cosmic strings for $G\mu = 10^{-12}$ in big dashed line.}\label{fig:gws}
\end{figure}
%%%%%%%%%%%%%%%%%%%%%%%%%%%%%%%%%%%%%%%%%%%%%%%%%%%%%%%%%%%%

Fig.~\ref{fig:gws} shows the gravitational wave background for the domain wall formation scale $\left<16_H\right>$  varying from $10^2$ to $10^5$ GeV, with  $G\mu = [10^{-15},10^{-9}]$, which corresponds to the breaking scales $\left<126_H\right>\sim [10^{11},10^{14}]$ GeV. The gravitational wave background will be detected \cite{Thrane:2013oya, Schmitz:2020syl} in several proposed experiments namely, CE \cite{PhysRevLett.118.151105}, ET \cite{Mentasti:2020yyd}, LISA \cite{Bartolo:2016ami, amaroseoane2017laser}, DECIGO \cite{Sato_2017}, BBO \cite{Crowder:2005nr, Corbin:2005ny} and HLVK \cite{KAGRA:2013rdx}, for $G\mu$ values from $10^{-15}$ to $10^{-9}$. For comparison we show the gravitational wave spectrum from topologically stable cosmic strings for $G\mu=10^{-12}$ in big dashed line. We can see that the lower frequency regions in the gravitational wave spectra from walls bounded by strings are absent since large loops with $l>2\pi R_c$ cannot be formed.  The NANOGrav \cite{NANOGrav:2023gor,NANOGrav:2023hvm} and other pulsar timing array experiments \cite{Antoniadis:2023ott,Reardon:2023gzh,Xu:2023wog} have recently reported evidence of stochastic gravitational wave background in the nano-Hertz frequency range. A recent study \cite{Lazarides:2023ksx} has shown that the gravitational waves from superheavy strings ($G\mu \approx 10^{-6}$) bounding the domain walls associated with a symmetry breaking scale in the ballpark of $10^2$ GeV can explain the data. It is worth mentioning that specific breaking chains of pGDM GUT models can give rise to similar topological structures and be compatible with the PTA data.

%%%%%%%%%%%%%%%%%%%%%%%%%%%%%%%%%%%%%%%%%%%%%%%%%%%%%%%%%%%%%%%%%%%
\section{Conclusions}
\label{sec:summary}
Pseudo-Goldstone dark matter (PGDM) models based on realistic grand unified gauge symmetries such as $SO(10)$ predict the existence of composite topological structures known as `walls bounded by strings'. In the $SO(10)$ model that we explore here, there exist a $126$-plet as well as $16$-plet of Higgs fields with suitable VEVs in order to obtain this dark matter particle. The $126$-plet VEV at an intermediate scale yields a topologically stable $\mathbb{Z}_2$ string, which then forms the boundary of the domain wall created when the $16$-plet acquires a VEV in the $10^2$-$10^5$ GeV range. We display the gravitational wave spectrum produced by this string-wall system and show that it will be accessible in the foreseeable future at a number of proposed experiments. We also point out that the gravitational wave emission from the topological structures discussed here with a string tension $G\mu$ close to $10^{-6}$ is compatible with the recently released NANOGrav 15 year data. 
%Finally, we note that the string-wall topological structure is expected to be present in other realistic models of pGDM including those based on a $U(1)_{B-L}$ extension of the Standard Model.

%%%%%%%%%%%%%%%%%%%%%%%%%%%%%%%%%%%%%%%%%%%%%%%%%%%%%%%%%%%%%%%%%%
\section{Acknowledgments}
R.M. acknowledges illuminating discussion with Joydeep Chakrabortty.  Q.S. thanks George Lazarides and Amit Tiwari for discussions on walls bounded by strings and related topological structures. 
This work was supported by research funds of Jeonbuk National University in 2022 (W.I.P.), by the National Research Foundation of Korea grants
by the Korea government: 2022R1A4A5030362 (R.M. and W.I.P.) and  2017R1D1A1B06035959 (W.I.P.), and by the Spanish grants PID2020-113775GB-I00 (AEI/10.13039/501100011033) and CIPROM/2021/054 (Generalitat Valenciana) (W.I.P.).

%%%%%%%%%%%%%%%%%%%%%%%%%%
\bibliographystyle{mystyle}
\bibliography{GUT_TD}
%%%%%%%%%%%%%%%%%%%%%%%%%%

\end{document}